\documentclass[aps,prb,twocolumn,amsfonts]{revtex4-2}
\usepackage{graphicx}
\usepackage{graphics}
\usepackage{xcolor}
\usepackage{amsmath}
\usepackage{xfrac}
\usepackage{natbib}
\usepackage[english]{babel}
\usepackage[colorlinks=true,citecolor=blue,linkcolor=blue,urlcolor=blue]{hyperref}
\usepackage{siunitx}

\begin{document}

\title{Statistically Modeling Optical Linewidths of Nitrogen Vacancy Centers in Microstructures}%

\author{M.~Kasperczyk, J.~A.~Zuber, A.~Barfuss, J.~K\"{o}lbl, V.~Yurgens, S.~Fl{\aa}gan, T.~Jakubczyk, B.~Shields, R.~J.~Warburton, P.~Maletinsky}%
\email{patrick.maletinsky@unibas.ch}
\affiliation{Department of Physics, University of Basel, Klingelbergstrasse 82, 4056 Basel, Switzerland}
\date{\today}%

\begin{abstract}
We investigate the relationship between ion implantation and the optical linewidth of the nitrogen vacancy (NV) zero-phonon line (ZPL) in bulk and structured samples.
We also propose a novel approach to ion implantation that we name post-implantation, in which nitrogen is implanted \textit{after} all fabrication processes have been completed.
We examine three post-implanted samples, one implanted with $^{14}$N and two with $^{15}$N isotopes.
We perform photoluminescence excitation (PLE) spectroscopy to assess optical linewidths and optically detected magnetic resonance (ODMR) measurements to isotopically classify the NV centers.
From this, we find that NV centers formed from nitrogen naturally occuring in the diamond lattice are characterized by a linewidth distribution peaked at an optical linewidth nearly two orders of magnitude smaller than the distribution characterizing most of the NV centers formed from implanted nitrogen.
Surprisingly, we also observe a number of $^{15}$NV centers with narrow ($<500\,\mathrm{MHz}$) linewidths, implying that implanted nitrogen can yield NV centers with narrow optical linewidths.
We further use a Bayesian approach to statistically model the linewidth distributions, to accurately quantify the uncertainty of fit parameters in our model, and to predict future linewidths within a particular sample.
Our model is designed to aid comparisons between samples and research groups, in order to determine the best methods of achieving narrow NV linewidths in structured samples.
\end{abstract}

\maketitle

\section{Introduction}

Excellent spectral properties and low spectral noise are a necessity for most quantum communications and entanglement protocols.
Whether the goal is to entangle atoms in different cities\,\cite{hofmann_heralded_2012}, to relay quantum information across vast distances through a communications channel\,\cite{yin_satellite-based_2017}, to couple a qubit to a photonic cavity\,\cite{faraon_coupling_2012,riedel_deterministic_2017}, or to study the interference between two qubits\,\cite{bernien_two-photon_2012}, some of the biggest successes of quantum technology rely on quantum sources that are spectrally stable\,\cite{batalov_temporal_2008}.
The nitrogen vacancy (NV) center in diamond has been particularly successful in a variety of quantum information experiments, as the NV spin can be coupled to its optical degree of freedom\,\cite{bernien_two-photon_2012, batalov_temporal_2008, faraon_coupling_2012, mouradian_scalable_2015, sipahigil_quantum_2012}.
Yet a key challenge remains in creating NV centers with good spectral properties in nano-structured samples, even though many groups have studied diverse methods of creating NVs.
These methods include implantation and annealing\,\cite{orwa_engineering_2011,fu_conversion_2010,chu_coherent_2014}, laser writing of NV centers\,\cite{chen_laser_2017,hadden_integrated_2018}, and high-energy electron irradiation\,\cite{ruf_optically_2019}, with many studies focusing explicitly on the linewidth properties of the NV centers\,\cite{tamarat_stark_2006, fu_observation_2009, robledo_control_2010, santori_nanophotonics_2010, siyushev_optically_2013}.
The transform limited optical linewidth of the NV center is $\approx\!13\,\mathrm{MHz}$, which sets the ultimate limit to how narrow the lines can be\,\cite{chu_coherent_2014}.
Depending on the application, broader optical linewidths can be tolerated: a $100\,\mathrm{MHz}$ linewidth is acceptable for a decent microcavity\,\cite{riedel_deterministic_2017}, and two-photon interference has been shown using an NV center with an inhomogeneous linewidth as broad as $480\,\mathrm{MHz}$\,\cite{bernien_two-photon_2012}.

Here we study the distribution of optical linewidths of NV centers formed with implanted and native nitrogen in diamond nanostructures.
We implant one of our samples with $^{15}$N, which has a natural abundance of only $0.37\%$\,\cite{rabeau_implantation_2006}, so that we can distinguish between implanted and native nitrogen by measuring the nitrogen isotope of the NV center.
In line with the results of S.~B. van~Dam \textit{et al.}\,\cite{van_dam_optical_2019}, we find that implanted nitrogen yields NV centers with generally broader linewidths than native nitrogen does.
These results improve our understanding of well-established and reliable fabrication recipes such as Chu \textit{et al.}\,\cite{chu_coherent_2014} by illuminating which types of NV centers are actually responsible for the narrow linewidths achieved--a consideration that was not evaluated in detail in those recipes.
We also find evidence that implanted nitrogen can yield NV centers with narrow linewidths.
Additionally, we propose the novel approach of post-implantation, in which all nano-structuring and fabrication procedures are completed \textit{before} implanting the sample with nitrogen.
We do this to reduce the effects of fabrication on the NV center properties, as it is unclear to what degree the fabrication procedures themselves influence the optical linewidth\,\cite{riedel_deterministic_2017,ruf_optically_2019}.
In studying post-implanted samples, we find a significant proportion of narrow linewidth NV centers, even in structured areas as thin as \SI{1.57}{\micro \metre}.
Finally, as determining what influences the NV center coherence properties remains an open question that is actively being explored, we develop a rigourous statistical model to help unify approaches within the community and to more easily compare results across research groups.
We discuss our model in depth and show how we can use it to compare different data sets.
To aid other researchers in implementing our model, we include a demo Matlab script, available as Supplementary Online Material\,\cite{SOM}.

\section{Overview of Samples}
\subsection{Fabrication Processes}

In the experimental part of this work, we first study two samples in detail (the third sample is discussed in Sec.~\ref{Sample_C}).
Both are made from electronic grade ($\mathrm{N}\,<5\,\mathrm{ppb}$, $\mathrm{B}\,<1\,\mathrm{ppb}$) diamond acquired from Element Six.
Our fabrication procedure is summarized by P.~Appel \textit{et al.}\,\cite{appel_fabrication_2016}.
In both samples, we fabricated a membrane of a nonuniform thickness spanning 2.5--\SI{5}{\micro \metre}, as well as cantilevers with variable dimension: lengths from 35--\SI{70}{\micro \metre}, widths of approximately \SI{4.5}{\micro \metre}, and thickness of roughly 2.5--\SI{4}{\micro \metre}.
An optical microscope image of Sample B is shown in Fig.~\ref{Fig1}(a), showing the cantilevers, membrane, and bulk parts of the sample.

\subsection{Implantation Parameters}

After all fabrication of the membrane and cantilevers was finished, Sample A was sent to the Helmholtz-Zentrum Dresden-Rossendorf to be implanted with nitrogen, and Sample B was sent to CuttingEdge Ions.
Both samples were implanted with $12\,\mathrm{keV}$ nitrogen ions at an angle of $7^\circ$ relative to the sample mount and at a fluence of $10^{11}\,\mathrm{ions/cm}^2$.
Whereas Sample A was implanted with $^{14}$N, Sample B was implanted with $^{15}$N, so that the NVs could be isotopically classified.
After the samples were implanted, they were annealed with a procedure outlined in P. Appel \textit{et al.}\,\cite{appel_fabrication_2016}, consisting of 4\,hours at 400\,$\mathrm{C}^\circ$, 10\,hours at 800\,$\mathrm{C}^\circ$, and 2\,hours at 1200\,$\mathrm{C}^\circ$.
Finally, the samples were cleaned with a tri-acid clean\,\cite{appel_fabrication_2016,brown_cleaning_2019}.

\section{Measurement Methodology}

\begin{figure}[tb]
	\centering
		\includegraphics[width=\columnwidth]{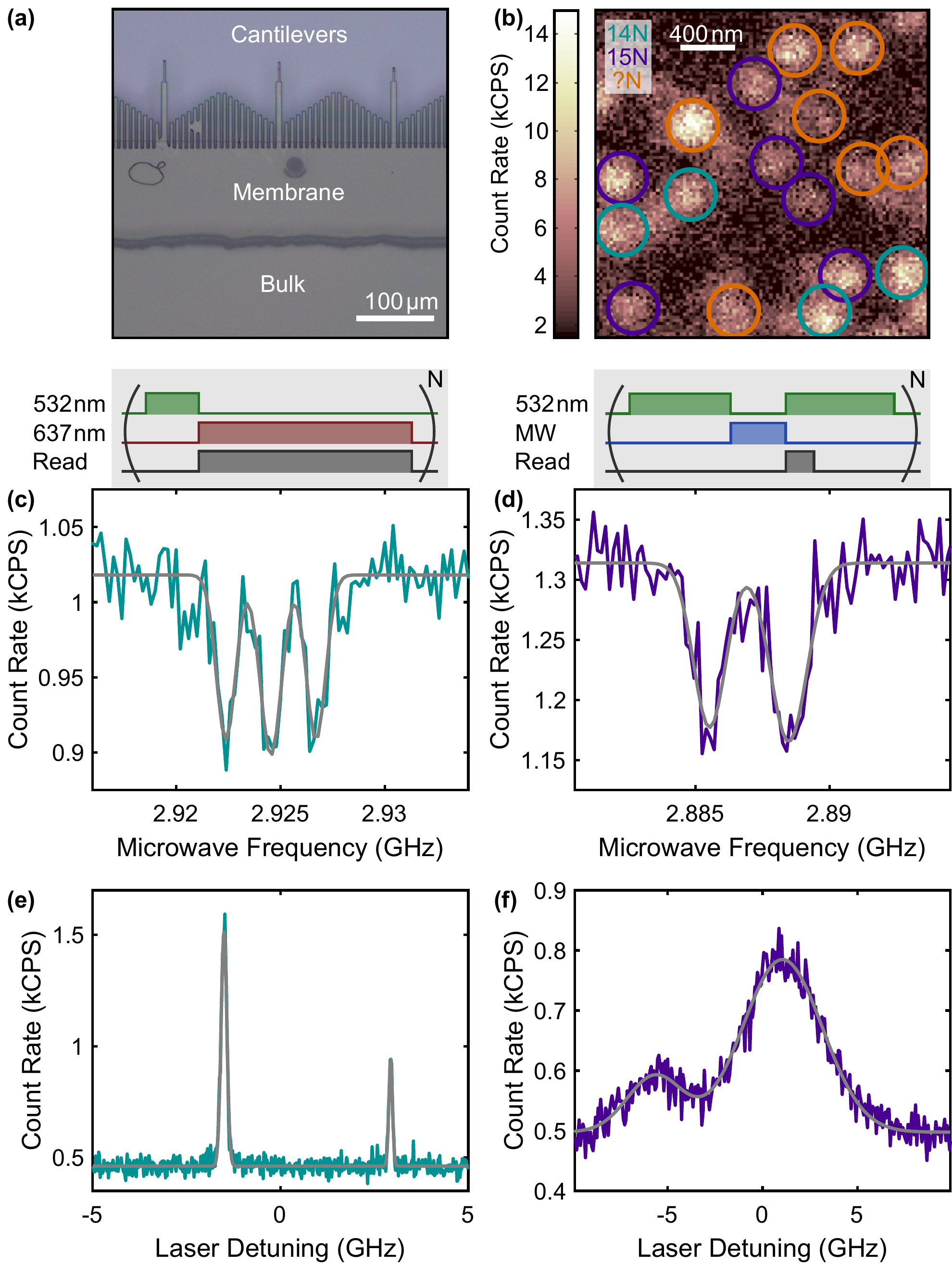}
	\caption{(a) Optical microscope image of Sample B, showing the bulk area (bottom), membrane (center), and cantilevers (top). (b) Fluorescence map in the bulk area of Sample B (\SI{50}{\micro \metre} thickness). Each NV center we measured is indicated by a circle around it, with the color of the circle indicating the isotope. Some NV centers indicated did not exhibit measurable PLE. (c) Representative PLE measurement on an $^{14}$NV, with the measurement pulse sequence shown above. The two lines have linewidths of $107\,\mathrm{MHz}$ and $186\,\mathrm{MHz}$ and are split by $4.4\,\mathrm{GHz}$. (d) Representative pulsed ODMR measurement on an $^{14}$NV center, with the pulse sequence shown above. Pulsed ODMR reveals the three peaks split by $2.2\,\mathrm{MHz}$ characteristic of $^{14}$N. (e) Representative PLE measurement on an $^{15}$NV center. The two lines have linewidths of $3.4\,\mathrm{GHz}$ and $5.1\,\mathrm{GHz}$ and are split by $6.8\,\mathrm{GHz}$. (f) Representative pulsed ODMR measurement on an $^{15}$NV center. Pulsed ODMR reveals the two peaks split by $3.1\,\mathrm{MHz}$ characteristic of $^{15}$N. In (c)--(f), the gray lines indicate fits to the data.}
	\label{Fig1}
\end{figure}

We begin by taking a confocal fluorescence map in the target area of the sample (see Fig.~\ref{Fig1}(b)).
We then characterize each potential NV center by taking a photoluminescence spectrum under green ($532\,\mathrm{nm}$) laser excitation.
Once the zero-phonon line (ZPL) has been identified on the spectrometer, we perform a photoluminescence excitation (PLE) measurement on the NV center by sweeping the wavelength of a red ($637\,\mathrm{nm}$) laser across the transition while recording the fluorescence counts on an avalanche photodiode (APD), yielding a measurement of the excited state transition linewidth.
We note that if the ZPL measured on the spectrometer had a linewidth above the resolution of our spectrometer (approximately \SI{70}{GHz} at \SI{637}{nm}), we did not attempt to measure PLE, and those measurements are not considered in the data sets we discuss later.
We did not use the same intensity of red laser power for each NV center, as the broader linewidths were often too weak to measure at low laser power.
Red laser powers for NV centers with narrow linewidths ranged from $10-200\,\mathrm{nW}$, whereas broad linewidths were typically measured with \SI{2}{\micro W} of excitation power. 
Optical linewidths are extracted from the FWHM of a Gaussian fit to the PLE data.
Because we include a repump pulse in every iteration of the pulse sequence, our linewidths are broadened by spectral diffusion, making a Gaussian fit suitable\,\cite{fu_conversion_2010,van_dam_optical_2019}.
The pulse sequence we use is shown above Fig.~\ref{Fig1}(c), and representative measurements for $^{14}$NV and $^{15}$NV centers are shown in Figs.~\ref{Fig1}(c)\&(e).
In the case in which two lines were visible, we inferred these to be the $E_x$ and $E_y$ lines, and we used only the narrower linewidth in the dataset, as the goal of this analysis is to analyze the narrowest linewidth measurable on each individual NV center.

After recording the optical linewidth, we use pulsed optical detection of magnetic resonance (ODMR) to measure the hyperfine structure of the NV center ground state, thereby identifying the isotope of the NV nitrogen.
The pulse sequence is shown above Fig.~\ref{Fig1}(d), and typical hyperfine-resolved ODMR measurements are shown in Figs.~\ref{Fig1}(d)\&(f) for $^{14}$N and $^{15}$N, respectively.
The locations and widths of the ODMR dips are extracted from Gaussian fits to the data.
We attempted to measure the ZPL wavelengths, optical linewidths, and hyperfine-resolved ODMR of a total of 159\,NV centers in Sample A and 104\,NV centers in Sample B.
We note, however, that some NV centers did not exhibit any PLE, and others failed to show hyperfine-resolved ODMR.
In total, we successfully measured PLE on 78\,NV centers in Sample A and 61\,NV centers in Sample B.
Similarly, we were able to isotopically classify 47\,NV centers on Sample B (isotopic classifcation was not performed on Sample A, as it was implanted with $^{14}$N).

\section{Results and Discussion}

\subsection{Influence of NV Center Location on Linewidth}

\begin{figure}[tb!]
	\centering
		\includegraphics[width=\columnwidth]{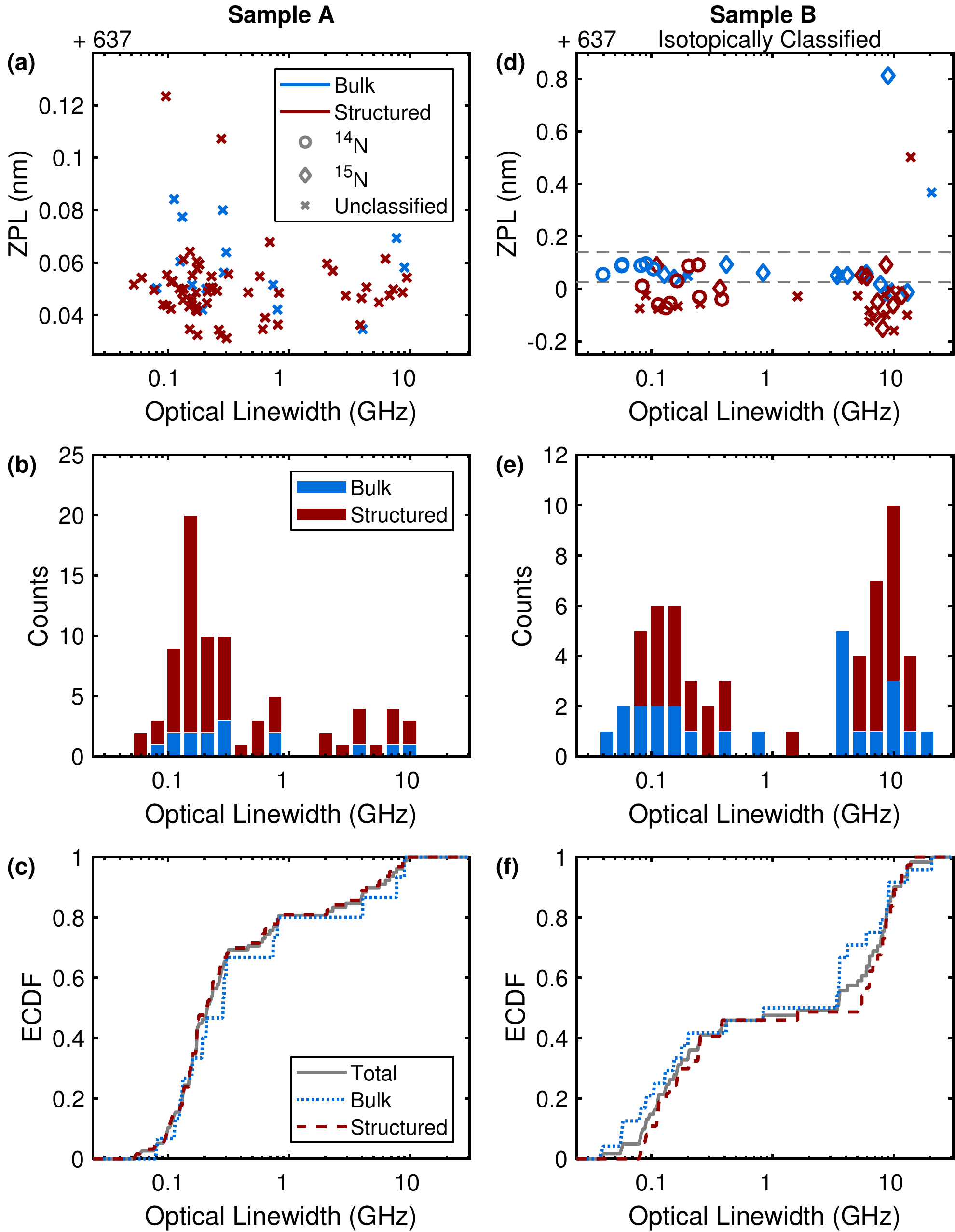}
	\caption{The data for Sample A are shown in the left column, and for Sample B in the right column. (a) A scatter plot showing the ZPL wavelength (in air) for each linewidth measured in Sample A. The marker color indicates which area of the sample the datapoint was taken on. The ZPL wavelength was measured on a spectrometer with a resolution of $0.9\,\mathrm{nm}$ (corresponding to $70\,\mathrm{GHz}$ resolution at $637\,\mathrm{nm}$), whereas the linewidth was measured using PLE. (b) Stacked histograms of optical linewidths in Sample A with the data labelled by sample location. There is evidence of two distinct populations of NV centers, and narrow NV centers can occur in the structured parts of the sample as well as the unstructured. (c) ECDFs of the linewidths in Sample A. The plot shows that the median measured linewidth was $\approx\,200\,\mathrm{MHz}$. (d) A similar scatter plot as in (a) for Sample B. The marker color again indicates where the datapoint was taken, and the marker shape indicates which isotope the hyperfine structure indicated. The horizontal dashed lines demarcate the limits of the ZPL axis of (a), showing that Sample B showed a much larger variation in ZPL wavelength, suggesting more variability in the local strain environment. (e) Stacked histogram of optical linewidths in Sample B, with data labelled by sample location. Two populations are again evident, and they are not related to location in the sample. (f) ECDF for Sample B. The median linewidth is $\approx 3.5\,\mathrm{GHz}$. The plateau indicates a clear separation between the two populations.}
	\label{Fig2}
\end{figure}

We summarize the data for Sample A in Fig.~\ref{Fig2}(a), which plots the measured optical linewidths against the measured ZPL wavelength.
The data points are color-coded to indicate which part of the sample they were taken on.
The bulk part of the sample is approximately \SI{50}{\micro \metre} thick, whereas the membrane and cantilever dimensions are discussed earlier. According to Wilcoxon ranked sum tests\,\cite{gelman_bayesian_2014}, the linewidths of NV centers found in the membrane likely follow the same statistics as those in the cantilevers. We therefore combine the cantilever and membrane measurements into a single category: structured.
The ZPL wavelengths for Sample A are tightly clustered (spanning a spectral range of only $0.2\,\mathrm{nm}$), and the sample exhibits no clear relationship between ZPL wavelength and optical linewidth.
Binning the linewidths and color-coding them according to the sample location (see Fig.~\ref{Fig2}(b)) reveals that there are two distinct populations of NV centers: those with narrow linewidths, and those with broad linewidths.
Both types of linewidths can be found anywhere on the sample.
We note that because the bins are plotted on a log scale, the bin widths are not constant, and bins at higher linewidths also cover a broader range of optical linewidths.
The widths of the bins in the plot therefore accurately represent their wavelength spans on the log axis.
We plotted the data in this way to be able to directly compare the narrow and broad linewidth distributions on the same plot while still achieving an appropriate resolution for all orders of magnitude within the data set.
Fig.~\ref{Fig2}(c) shows the empirical cumulative distribution functions (ECDFs) for the structured, bulk, and total datasets for Sample A, showing that there is no apparent difference between the three distributions.
A Wilcoxon ranked sum test ($p$-value $0.551$) reveals that there is no statistically significant evidence that the structured and bulk linewidth distributions are different.
Fig.~\ref{Fig2}(c) also shows that the median measured linewidth was approximately $200\,\mathrm{MHz}$.

In Figs.~\ref{Fig2}(d)--(f), we show similar plots for Sample B.
In Fig.~\ref{Fig2}(d) we see that although the ZPL wavelengths are far more scattered in Sample B (spanning a range of $2\,\mathrm{nm}$) than in Sample A, there is still no clear relationship between ZPL wavelength and optical linewidth, indicating that local strain does not play a strong role in determining the linewidth.
Plotting the data in a histogram labelled by location of the sample in Fig.~\ref{Fig2}(e) shows a similar situation as in Fig.~\ref{Fig2}(b): there are two distinct populations of NV centers, independent of the location on the sample.
Likewise, Fig.~\ref{Fig2}(f) leads to similar conclusions as Fig.~\ref{Fig2}(c).
Again, a Wilcoxon rank sum test ($p$-value $0.334$) indicates that there is no clear evidence for a difference between the linewidth distributions in the bulk and structured areas in Sample B.
As we see similar results in two different samples, and across different regions on those samples, we turn to isotopic classification to better understand these two populations.

\subsection{Influence of Ion Implantation on Linewidth}

\begin{figure}[tb]
	\centering
		\includegraphics[width=\columnwidth]{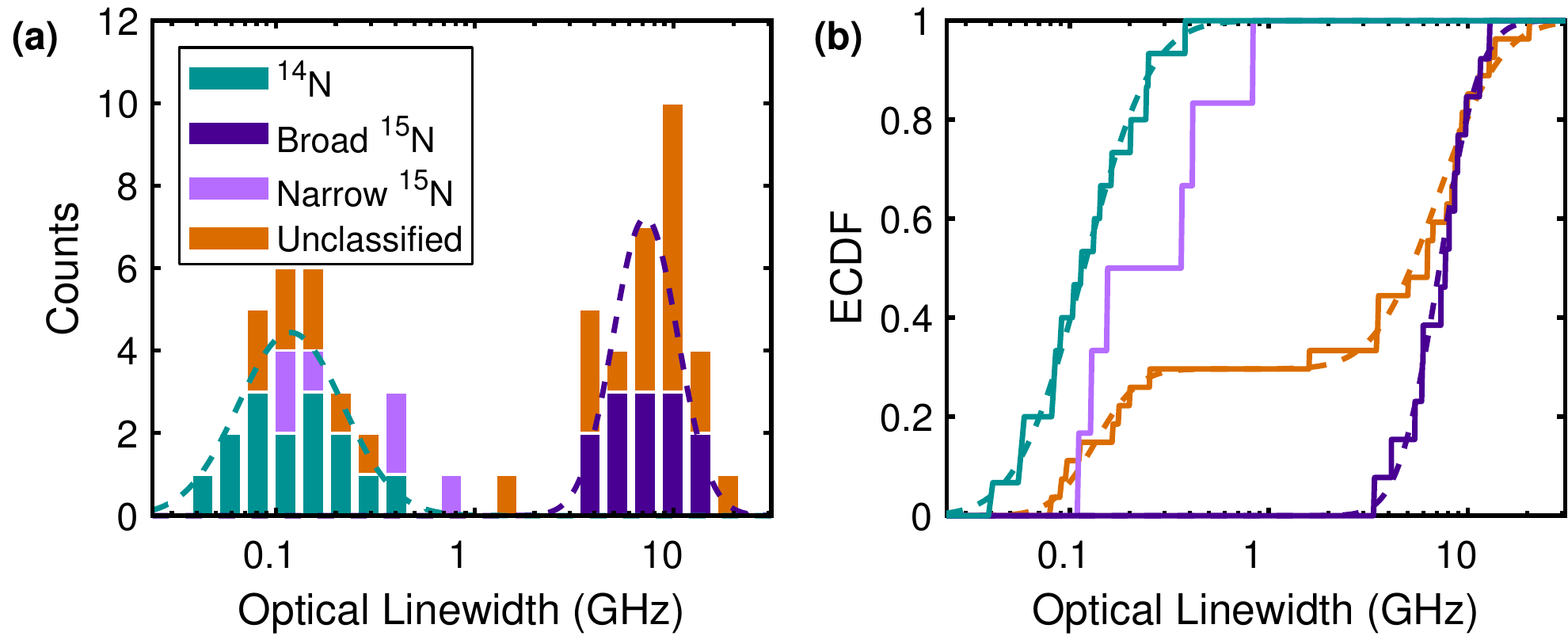}
	\caption{(a) Stacked histogram of optical linewidths for Sample B, with data points labelled by color to indicate isotope. The histogram clearly shows that, with a few exceptions, the $^{14}$NV centers and the $^{15}$NV centers are completely separated into two distinct populations. The dashed lines indicate fits of log-normal sampling distributions $P(\lbrace x_i \rbrace \vert \mu, \sigma^2)$ to the $^{14}$NV and (broad) $^{15}$NV datasets. (b) Isotopically classified ECDFs. The teal and purple dashed lines are the log-normal fits to the $^{14}$NV and $^{15}$NV datasets, respectively, showing excellent agreement between the ECDFs and the fits. The dashed orange line is obtained by fitting a sum of two log-normal distributions to the unclassified NV centers. The narrow $^{15}$NV data are included without a fit in this plot.}
	\label{Fig3}
\end{figure}

In Fig.~\ref{Fig3}(a) we bin the data and color-code the bins by isotope classification.
We find that although many NV centers could not be clearly classified, a clear pattern emerges: NV centers formed with native $^{14}$N exhibit narrow ($<1\,\mathrm{GHz}$) linewidths, whereas most of the $^{15}$NV centers showed broad ($>1\,\mathrm{GHz}$) linewidths, in agreement with the results of S.~B. van~Dam\,\cite{van_dam_optical_2019}.
Indeed, the median $^{14}$NV linewidth in Sample B was roughly $100\,\mathrm{MHz}$.
We fit log-normal sampling distributions to the $^{14}$NV and $^{15}$NV data (dashed lines).
In Fig.~\ref{Fig3}(b) we plot the ECDFs for the isotopically classified datasets, as well as the cumulative distribution functions (CDFs) for the log-normal fits.
The CDFs show exceptional agreement with the ECDFs (see Sec.~\ref{Diagnostics} for model diagnostics).
The fit curve for the unclassified data set comprises a weighted sum of two log-normal distributions in which the weights are also fit parameters. This fit suggests that all the unclassified data can be attributed to one of the two distributions.

Finally, we note that six of the $^{15}$NV center linewidths were well separated from those of the other $^{15}$NV centers.
Based on an analysis of quantile-quantile (Q-Q) plots (see the discussion of Q-Q plots in Sec.~\ref{Diagnostics}), we exclude these NV centers from the fits in Fig.~\ref{Fig3}(a), as they clearly do not belong to the same population; in Fig.~\ref{Fig3}(b) we include the ECDF of these data points but do not fit them.
Due to the low natural abundance of $^{15}$N, it is highly unlikely that the narrow $^{15}$NV centers are due to naturally occurring $^{15}$N.
To wit: in a sample size of 61 PLE lines, there is a mere $1.2 \times 10^{-5}\%$ chance of observing 6 or more naturally occurring $^{15}$NV centers, i.e. $P\left(m \geq 6 \vert n = 61, p = 0.0037\right) \approx 1.2 \times 10^{-7}$, calculated from the CDF of the binomial distribution with 61 trials and a success rate of 0.37\%.
Previous studies have reported that implanted nitrogen can lead to crystal damage that degrades the optical properties of NV centers, and that this damage can be at least partially repaired through annealing\,\cite{van_dam_optical_2019}, but it is unclear whether the annealing is the reason we were able to observe narrow linewidths from NV centers formed by implanted nitrogen.

\section{Statistical Model}

\subsection{Building the Model}

We now develop a model to describe the two distinct populations we see, as it could be useful to determine how different the populations are. A model could help to decide how we should classify future or unclassified data points, and to predict how narrow future linewidths in the same sample will be.
Additionally, having a model will allow us to more quantitatively determine which fabrication procedures yield NV centers with better optical linewidths and quantify how certain we are a new procedure is better.
Using a Bayesian approach, we model the likelihood of a particular linewidth $x_i$ with a log-normal likelihood:

\begin{align}
P\left( x_i \vert \mu, \sigma \right) = \frac{1}{\sqrt{2 \pi \sigma}} \frac{1}{x_i} e^{-\sfrac{\left( \mathrm{ln}\! \left( x_i \right) - \mu \right)^2}{2 \sigma^2}} \, , \label{Eq1}
\end{align}

\noindent which is parameterized by a median $\mu$ and a standard deviation $\sigma$.
This is an appropriate distribution for any purely positive quantity that has contributions from multiple independent noise sources\,\cite{sivia_data_2006} (here, e.g., electric field noise, temperature, and strain fluctuations can all influence the optical linewidth\,\cite{tamarat_stark_2006,fu_observation_2009}).

Using uninformative priors for the parameters, (uniform distribution for $\mu$ and the Jeffreys prior for $\sigma$\,\cite{sivia_data_2006}), we find their posterior distributions\,\cite{gelman_bayesian_2014,sivia_data_2006}.
See Appendix~\ref{AppA} for details.
Broadly speaking, the posterior distributions describe our best guess for the parameters, as well as our confidence in those guesses, given the data we have and the model we use.
For ease of notation, we define two constants that depend on the data:

\begin{align*}
\overline{X} &= \frac{1}{N} \sum^N_{i=1}{\left( \mathrm{ln} \left( x_i \right) \right)} \, , \\
\overline{X^2} &= \frac{1}{N} \sum^N_{i=1}{\left( \mathrm{ln} \left( x_i \right) \right)^2}\, ,
\end{align*}

\noindent where $x_i$ is the $i$th linewidth in the dataset (or data subset, if focusing on a particular isotope, for example) and $N$ is the total number of linewidths in the dataset (or subset).

For $\mu$, we find that the posterior distribution $P(\mu \vert \lbrace x_i \rbrace)$ (where $\lbrace x_i \rbrace$ is the dataset of linewidths being analyzed) is given by a location-scale t-distribution:

\begin{align}
P&(\mu \vert \lbrace x_i \rbrace) = \nonumber \\ &\frac{\Gamma \left( \frac{\nu_\mu + 1}{2} \right)}{\Gamma \left( \frac{\nu_\mu}{2} \right) \sqrt{\pi \nu_\mu {\sigma_\mu}^2}} \left( 1 + \frac{1}{\nu_\mu} \left( \frac{\mu - \mu_\mu}{\sigma_\mu} \right)^2 \right)^{-\frac{\nu_\mu + 1}{2}}\, , \label{Eq2}
\end{align}

\noindent where $\mu_\mu = \overline{X}$, $\nu_\mu = N-1$, and $\sigma_\mu = \sqrt{\frac{1}{N-1} \left( \overline{X^2} - \overline{X}^2 \right)}$.

For the variance $\sigma^2$, the posterior distribution $P(\sigma^2 \lvert \lbrace x_i \rbrace)$ is an inverse gamma distribution:

\begin{align}
P(\sigma^2 \vert \lbrace x_i \rbrace) = \frac{{\beta_\sigma}^{\alpha_\sigma}}{\Gamma \left( \alpha_\sigma \right)} \left(\frac{1}{{\sigma}^2} \right)^{\left( \alpha_\sigma + 1 \right)} e^{-\sfrac{\beta_\sigma}{{\sigma}^2}} \, , \label{Eq3}
\end{align}

\noindent where $\alpha_\sigma = \frac{N-1}{2}$ and $\beta_\sigma = \frac{N}{2} \left( \overline{X^2} - \overline{X}^2 \right)$.

We next consider what distribution of future linewidths $\tilde{x}$ we expect to measure, given the data we have observed so far. Working in terms of the natural logarithm of the linewidth $\tilde{X} \equiv \mathrm{ln}\!\left( \tilde{x} \right)$, we also calculate the posterior predictive distribution $P(\tilde{X} \vert \lbrace x_i \rbrace)$, which describes how likely the next linewidth is to be narrow.
We find that $P(\tilde{X} \vert \lbrace x_i \rbrace)$ is a location-scale t-distribution:

\begin{align}
P&(\tilde{X} \vert \lbrace x_i \rbrace) = \nonumber \\ &\frac{\Gamma \left( \frac{\tilde{\nu} + 1}{2} \right)}{\Gamma \left( \frac{\tilde{\nu}}{2} \right) \sqrt{\pi \tilde{\nu} \tilde{\sigma}^2}} \left( 1 + \frac{1}{\tilde{\nu}} \left( \frac{ \tilde{X} - \tilde{\mu} }{\tilde{\sigma}} \right)^2 \right)^{- \frac{\tilde{\nu} + 1}{2}} \, , \label{Eq4}
\end{align}

\noindent where $\tilde{\mu} = \overline{X}$, $\tilde{\nu} = N-1$, and $\tilde{\sigma} = \sqrt{\frac{N+1}{N-1} \left( \overline{X^2} - \overline{X}^2 \right)}$.
Note that this is a location-scale t-distribution for the natural logarithm of the linewidth $\tilde{X}$, not for the linewidth $\tilde{x}$ itself.
The posterior predictive distribution for the linewidth $\tilde{x}$ is given by

\begin{align}
P&(\tilde{x} \vert \lbrace x_i \rbrace) = \nonumber \\ &\frac{\Gamma \left( \frac{\tilde{\nu} + 1}{2} \right)}{\Gamma \left( \frac{\tilde{\nu}}{2} \right) \sqrt{\pi \tilde{\nu} \tilde{\sigma}^2}} \frac{1}{\tilde{x}} \left( 1 + \frac{1}{\tilde{\nu}} \left( \frac{ \mathrm{ln} \! \left( \tilde{x} \right) - \tilde{\mu} }{\tilde{\sigma}} \right)^2 \right)^{- \frac{\tilde{\nu} + 1}{2}} \, , \label{Eq5}
\end{align}

\noindent which is not quite a t-distribution.
For a more detailed discussion of the derivations of these distributions and their interrelations, see Appendix~\ref{AppA}.

\subsection{Inferences from the Model}

Because these distributions are of a common form, it is straightforward to find their most likely values and their credible intervals.
For example, the maximum a posterior (MAP) estimate (i.e. the most likely value) for the t-distribution $P(\mu \vert \lbrace x_i \rbrace)$ is given by

\begin{align*}
\mu^{\mathrm{MAP}} = \mu_\mu = \overline{X} \, ,
\end{align*}

\noindent and the $95\%$ credible interval is given by

\begin{align*}
\left[ \mu_\mu - \sigma_\mu t_{(0.975, \nu_\mu)}, \mu_\mu + \sigma_\mu t_{(0.975, \nu_\mu)} \right] \, ,
\end{align*}

\noindent where $t_{(f,\nu)}$ is the $t$-statistic at the $f$th percentile and with $\nu$ degrees of freedom\,\cite{gelman_bayesian_2014,sivia_data_2006}.
Note, however, that the parameter $\mu$ in the log-normal distribution has units of $\mathrm{ln}\!\left( \mathrm{MHz} \right)$ (if the dataset is in MHz); the MAP estimate and credible interval (CI) in terms of $\mathrm{MHz}$ are then given by

\begin{align*}
&\qquad \quad \qquad \quad e^{\mu_\mu} \: \mathrm{and} \\ &\left[ e^{\mu_\mu - \sigma_\mu t_{(0.975, \nu_\mu)}}, e^{\mu_\mu + \sigma_\mu t_{(0.975, \nu_\mu)}} \right] \, ,
\end{align*}

\noindent respectively.
Similarly, the MAP estimate and CI for $\tilde{x}$ are given by

\begin{align*}
&\qquad \quad \qquad e^{\tilde{\mu}} \: \: \mathrm{and} \\ &\left[ e^{\tilde{\mu} - \tilde{\sigma} t_{(0.975, \tilde{\nu})}}, e^{\tilde{\mu} + \tilde{\sigma} t_{(0.975, \tilde{\nu})}} \right] \, ,
\end{align*}

\noindent respectively. 
The MAP estimate of $\sigma^2$ is given by $\sfrac{\beta_\sigma}{\left( \alpha_\sigma + 1 \right)}$\,\cite{gelman_bayesian_2014}.
Unfortunately, there is no closed-form solution for the 95\% CI of the inverse gamma distribution, but it can be easily estimated through simulated draws, which we describe below\,\cite{gelman_bayesian_2014}.

\begin{figure}[tb!]
	\centering
		\includegraphics[width=\columnwidth]{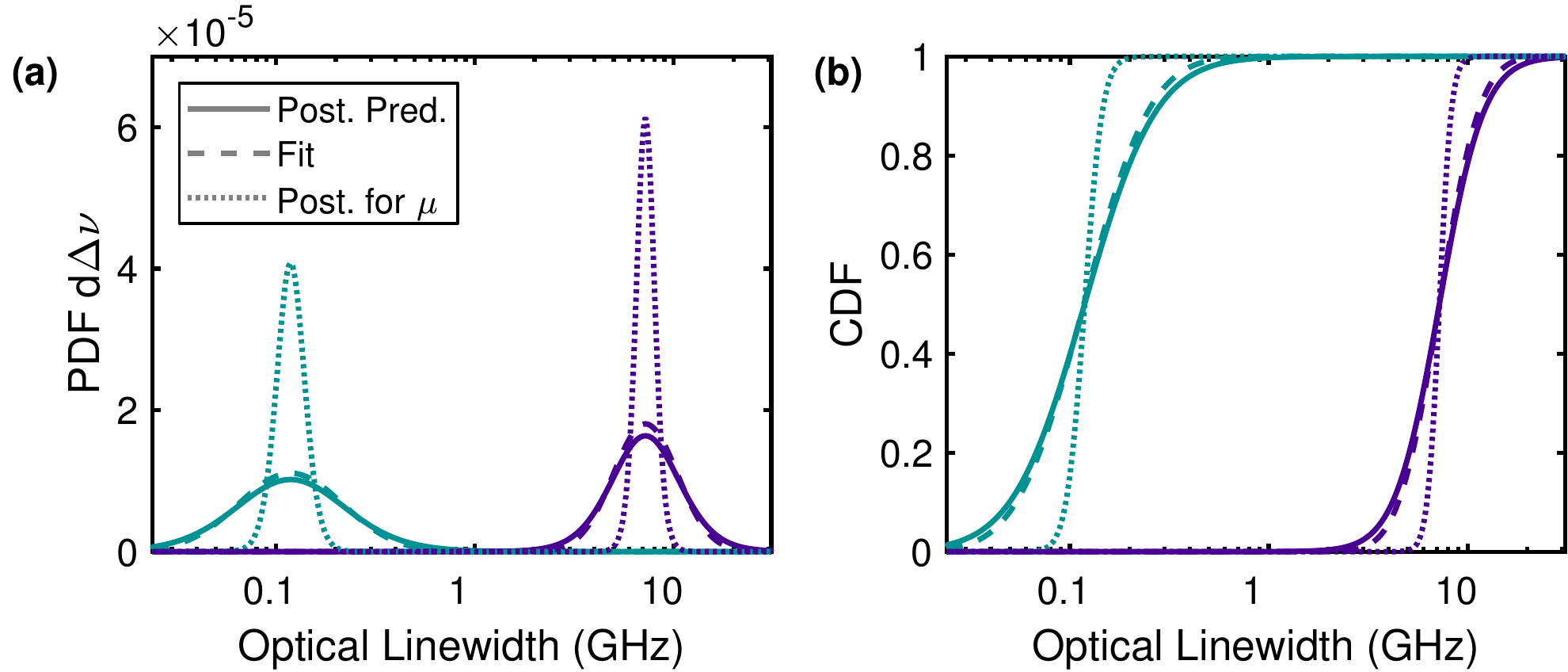}
	\caption{(a) The dashed lines reproduce the log-normal sampling distributions from Fig.~\ref{Fig3}(a). The solid lines indicate the posterior predictive distributions $P\left(\tilde{x} \vert \lbrace x_i \rbrace \right)$. They are slightly broader than the sampling distributions $P\left(x \vert \mu, \sigma\right)$ because they account for the uncertainty in our estimates of $\mu$ and $\sigma$. The dotted lines are the posterior distributions for the median $P(\mu \vert \lbrace x_i \rbrace)$, showing that the median $^{14}$NV and $^{15}$NV linewidths are well separated. (b) CDFs of the corresponding distributions in (a).}
	\label{Fig4}
\end{figure}

We graphically represent our results in Fig.~\ref{Fig4}.
The dashed lines in Fig.~\ref{Fig4}(a) are the log-normal fits from Fig.~\ref{Fig3}(a).
As in Fig.~\ref{Fig3}, the color of the line indicates the isotope.
The solid lines are the posterior predictive distributions $P \left( \tilde{x} \vert \lbrace x_i \rbrace \right)$, and the dotted lines are the posterior distributions $P \left( \mu \vert \lbrace x_i \rbrace \right)$.
The posterior predictive distributions $P \left( \tilde{x} \vert \lbrace x_i \rbrace \right)$ resemble the sampling distributions $P \left( \lbrace x_i \rbrace \vert \mu, \sigma \right)$ but are slightly broader, as they account for the uncertainty in our estimates of $\mu$ and $\sigma$.
The posterior for $\mu_{\mathrm{N14}}$ given by $P \left( \mu_{\mathrm{N14}} \vert \lbrace x_i \rbrace_{N14} \right)$ is fairly narrow, indicating that only a narrow range of values of $\mu_{\mathrm{N14}}$ is consistent with the $^{14}$NV data.
Similar conclusions hold for the $^{15}$NV data.

Finally, we simulate draws from the distributions, which allows us to compare the $^{14}$NV and $^{15}$NV results and give approximate answers to questions such as what is the probability that the next $^{14}$NV linewidth is narrower than the next $^{15}$NV linewidth $P\left(\tilde{x}_{\mathrm{N}14} < \tilde{x}_{\mathrm{N}15} \vert \lbrace x_i \rbrace \right)$ or how likely is the next $^{14}$NV linewidth to be below $100\,\mathrm{MHz}$ $P\left(\tilde{x}_{\mathrm{N}14} < 100\,\mathrm{MHz} \vert \lbrace x_i \rbrace \right)$.
For example, using our data and $10^8$ simulated draws from each of the posterior distributions, we find that $P\left( \mu_{\mathrm{N14}} < \mu_{\mathrm{N15}} \vert \lbrace x_i \rbrace \right) \approx 1$.
Similarly, we estimate that we have a roughly $40\%$ chance of finding sub-$100\,\mathrm{MHz}$ $^{14}$NV centers: $P \left( \tilde{x}_{\mathrm{N14}} < 100\,\mathrm{MHz} \vert \lbrace x_i \rbrace \right) \approx 0.398$.
For details of the simulated draws, see Appendix~\ref{AppB}.

There are a number of other tests we can perform and other questions we can answer. For example, are the narrow-linewidth NVs in Sample A characterized by the same median as the narrow-linewidth NVs in Sample B? We approach this question by splitting the data sets for both samples into narrow (\SI{<1000}{GHz}) and broad (\SI{\geq 1000}{GHz}) subsets. Using the same formulae as above, we can then evaluate $P\left( \mu_{\mathrm{narrow}}^{A} < \mu_{\mathrm{narrow}}^{B}  \right) \approx 0.982$. Similarly, if we compare the broad optical linewidths in both samples, we find $P\left( \mu_{\mathrm{broad}}^{A} < \mu_{\mathrm{broad}}^{B}  \right) \approx 0.021$. It is therefore unlikely (although not impossible) that the same medians apply in both samples. This could be explained by the fact that although the implantation parameters were nominally identical, the implantation was done by different companies and at different times, leading to slight but measurable difference in the NV properties between the two samples. On the other hand, we can compare the unclassified NVs in Sample B to the $^{14}\mathrm{NVs}$ and the broad $^{15}\mathrm{NVs}$. Again, we split the unclassified NVs into narrow and broad categories, with the boundary at \SI{1}{GHz}. We then find that $P \left( \mu_{\mathrm{N14}} < \mu_{\mathrm{Unc,narrow}}\right) \approx 0.712$, and $P \left( \mu_{\mathrm{N15}} < \mu_{\mathrm{Unc,broad}} \right) \approx 0.478$. It is thus reasonable to characterize the unclassified NVs by the same medians as the $^{14}\mathrm{NVs}$ and $^{15}\mathrm{NVs}$.

\subsection{Model Checking} \label{Diagnostics}

\begin{figure}[tb]
	\centering
		\includegraphics[width=\columnwidth]{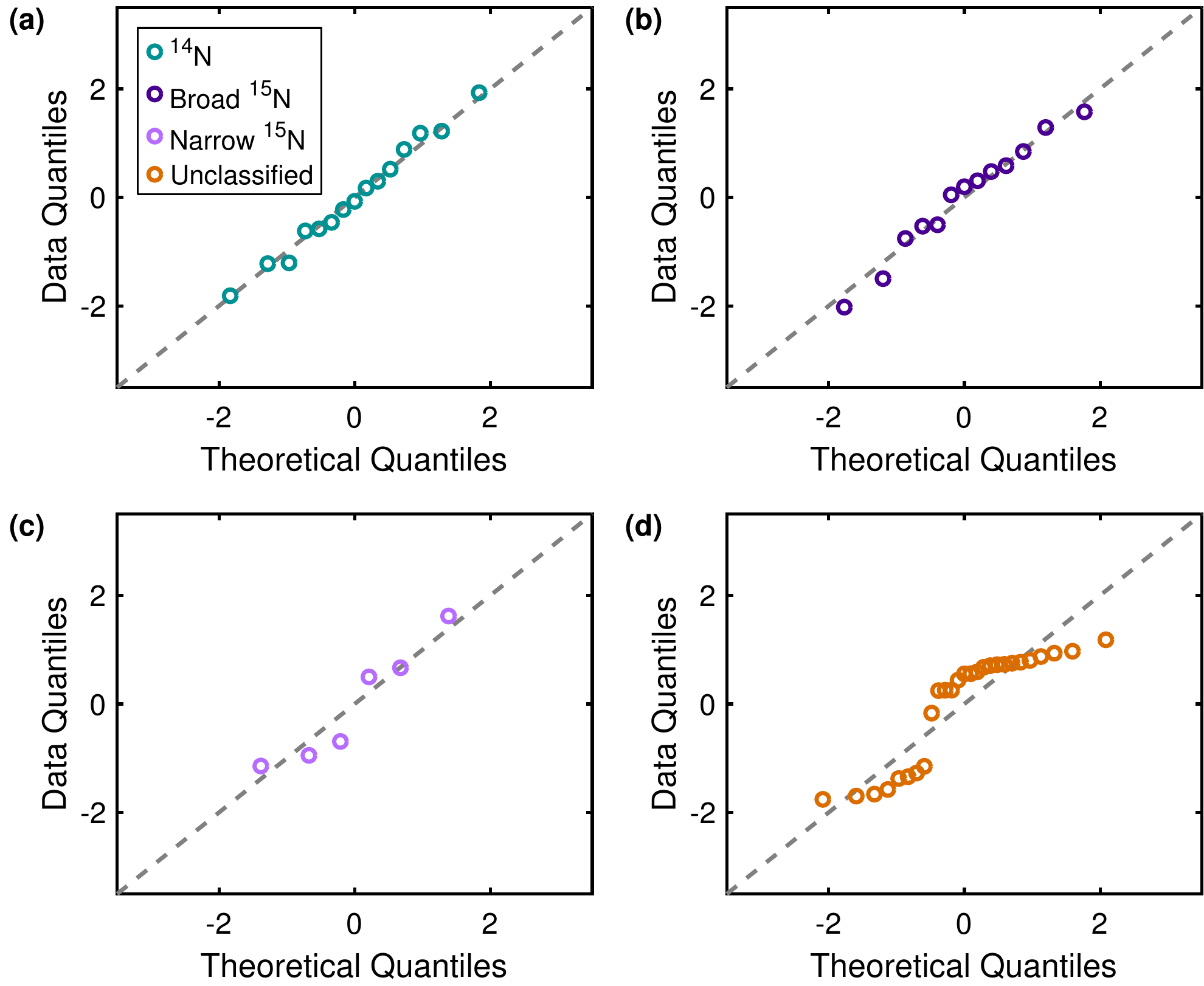}
	\caption{(a) Q-Q plot for the $^{14}$NV data, showing that a log-normal model fits the data well. (b) Q-Q plot for the $^{15}$NV data, again showing the model choice was appropriate. (c) The Q-Q plot for the narrow $^{15}$NV data shows that a log-normal model might be appropriate. (d) The Q-Q plot for the unclassified data shows that a single log-normal model is a poor fit to the data, as expected.}
	\label{Fig5}
\end{figure}

To check how appropriate our model is for our data, we look at the quantile-quantile (or Q-Q) plots for the different data subsets.
By comparing the data quantiles to the expected quantiles from the model, Q-Q plots show whether the spread in the data can be explained by the model and are therefore a useful diagnostic for determining whether a model is appropriate for the data.
They can also be useful for identifying outliers in the dataset.
The quantile for the $i$th optical linewidth in the dataset is calculated according to the formula

\begin{align*}
Q_{\mathrm{Data}}^{\left(i\right)} = \frac{\mathrm{ln} \! \left( x_i \right) - \overline{X}}{\sqrt{\overline{X^2} - \overline{X}^2}}
\end{align*}

\noindent and therefore summarizes how many standard deviations the data point is from the mean of the dataset\,\cite{thode_testing_2002}.
Using a log-normal model to calculate the theoretical quantiles, we plot the Q-Q plots for the $^{14}$NV data, the broad $^{15}$NV data, the narrow $^{15}$NV data, and the unclassified data in Fig.~\ref{Fig5}.
Both the $^{14}$NV data in Fig.~\ref{Fig5}(a) and the broad $^{15}$NV data in Fig.~\ref{Fig5}(b) closely follow the diagonal dashed line, indicating the quantiles of the measured data match the quantiles we would expect from a log-normal distribution in both cases.
Due to the dearth of data points, it is difficult to say how appropriate a log-normal model is for the narrow $^{15}$NV data in Fig.~\ref{Fig5}(c), but our data do show that a log-normal model is promising.
From Fig.~\ref{Fig5}(d), it is clear that a single log-normal model is inappropriate for the unclassified data, as expected.

\subsection{Example with Sample C} \label{Sample_C}

\begin{figure}[tb]
	\centering
		\includegraphics[width=\columnwidth]{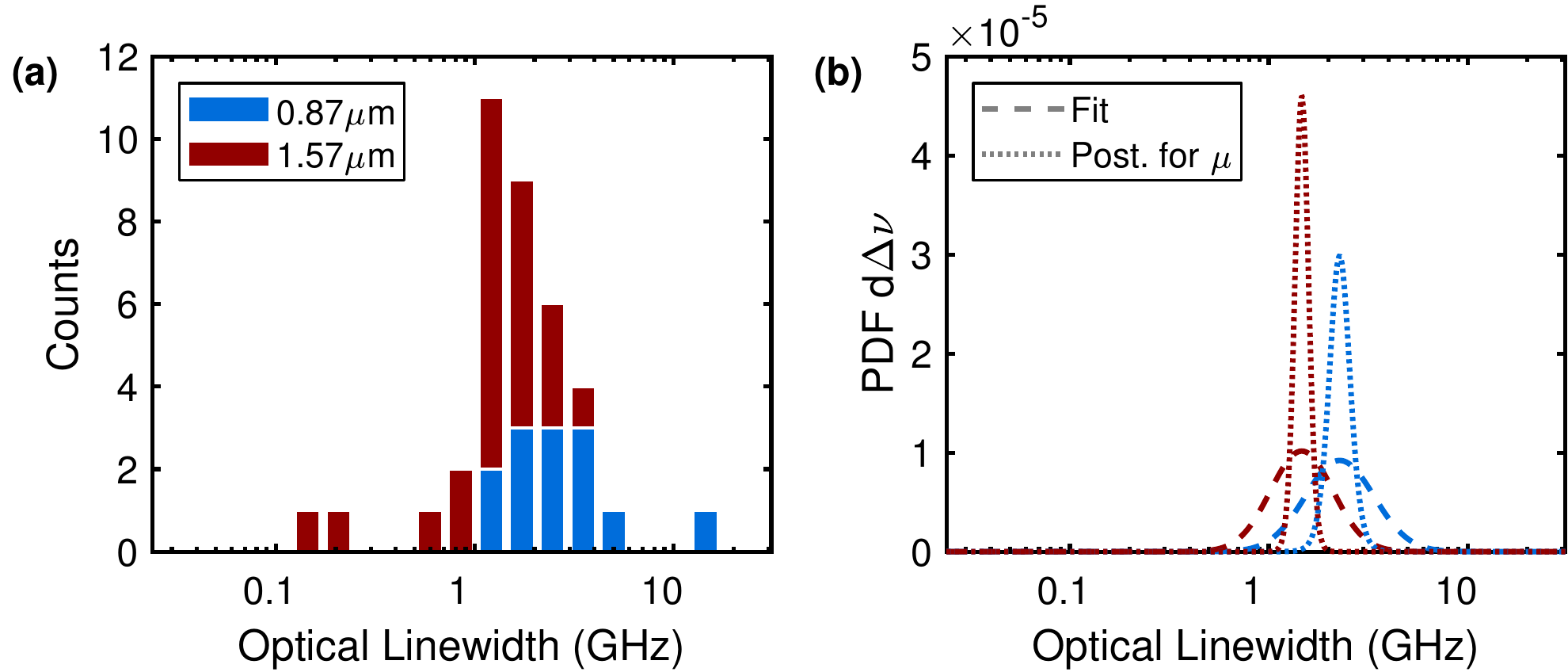}
	\caption{(a) Stacked histograms of data taken in Sample C. The data are color coded to indicate which area of the sample they were taken on. (b) Fits (dashed lines) and posterior distributions for $\mu$ (dotted lines) for the two sample areas. The plots are color coded as in (a). Although the fits strongly overlap, the posterior distributions for $\mu$ do not, indicating that the thinner part of the sample is characterized by a higher median linewidth.}
	\label{Fig6}
\end{figure}

As an application of our statistical model, we now examine a third structured sample, Sample C, which was post-implanted by InnovIon with $52\,\mathrm{keV}$ $^{15}$N ions at an angle of $7^\circ$ and a fluence of $5 \times 10^{9}\,\mathrm{ions/cm}^2$.
In Sample C, we compare two structured parts of the sample: one area that is \SI{1.57}{\micro \metre} thick, and one that is \SI{0.87}{\micro \metre} thick.
First, we note that we were able to observe two narrow ($<250\,\mathrm{MHz}$) linewidths in the \SI{1.57}{\micro \metre}-thick area (see Fig.~\ref{Fig6}(a)).
To our knowledge, these are the narrowest NV ZPL lines reported in such thin structures obtained by standard etching techniques.
A recent report, however, suggests that ultra-slow etching can significantly improve surface quality and lead to a further reduction of charge noise, which is at the origin of the inhomogenous broadening\,\cite{lekavicius_diamond_2019}.
We note that the distributions of the data from the two sample areas strongly overlap (see Fig.~\ref{Fig6}(a)).
In Fig.~\ref{Fig6}(b), we show log-normal fits to the data and the posterior distributions for $\mu$.
We find that the two data subsets have similar MAP estimates for the medians: $\mu_{\SI{0.87}{\micro \metre}}^\mathrm{MAP} \approx \mathrm{ln}\!\left( 2.27\,\mathrm{GHz} \right)$ and $\mu_{\SI{1.57}{\micro \metre}}^\mathrm{MAP} \approx \mathrm{ln}\!\left( 1.47\,\mathrm{GHz} \right)$.
Although the data and fits overlap and the estimates for $\mu$ are similar for the two data subsets, the posterior distributions for the medians $\mu$ barely overlap.
Using the data from Sample C and simulated draws from the posterior distributions for $\mu$, we find that $P\left( \mu_{\SI{1.57}{\micro \metre}} < \mu_{\SI{0.87}{\micro \metre}} \right) \approx 0.996$, strongly suggesting that the two areas have different median linewidths.
For the purposes of estimating the two medians, we exclude the two narrowest linewidths in the \SI{1.57}{\micro \metre} area of the sample and the broadest linewidth in the \SI{0.87}{\micro \metre} area of the sample, as Q-Q plots (not shown) reveal these data points to be outliers.
Although our data suggest that the thinner part of the sample has a larger median linewidth, it is unclear whether this change is due the thickness itself or due to confounding variables.
We hope our statistical model will aid in determining which variables influence the spectral properties of NV centers in other nano-structured samples.

\section{Conclusion}

We have shown that NV centers in post-implanted samples exhibit narrow linewidths, even in structured samples as thin as \SI{1.57}{\micro \metre}, and that the narrow lines are primarily due to NV centers being formed from nitrogen native to the diamond.
Even so, we observe a few narrow linewidths that can be attributed to implanted nitrogen.
Furthermore, we develop a statistical model to aid in summarizing our results and to enable easy comparison of results between research groups.
Indeed, we employ our model to show that in one of our samples, the sample thickness is linked to changes in the linewidth distribution.
Our results show that post-implantation is capable of yielding NV centers with narrow optical linewidths.
To further investigate the benefits of post-implantation, we propose testing different fabrication steps on a post-implanted structured sample, to study if and how various common fabrication techniques degrade the NV properties.
If post-implantation can be shown to improve NV coherence properties, it is worthwhile to study the effects of carbon implantation, as implanted nitrogen rarely leads to narrow linewidths.
Finally, the model itself can be developed further, by implementing a hierarchical model (to allow, e.g., isotope abundance to vary across the sample, or to allow the parameters $\mu$ and $\sigma$ to vary with sample location) and by including a model for data sampling and missing data.

\begin{acknowledgments}
We thank R.~Hanson and M.~Degen for fruitful discussions. We gratefully acknowledge financial support through the NCCR QSIT (Grant No. 185902), through Swiss NSF Project Grant Nos. 188521 and 155845, through the Innovative Training Network (ITN) SpinNANO, and through the EU Quantum Flagship project ASTERIQS (Grant No. 820394).
\end{acknowledgments}

\appendix

\section{Derivation of Posterior Probabilities} \label{AppA}
We find the joint posterior distribution for the parameters $\mu$ and $\sigma$ by using Bayes law:

\begin{align*}
P\left( \mu, \sigma \vert \lbrace x_i \rbrace \right) = \frac{P\left(\lbrace x_i \rbrace \vert \mu, \sigma \right) P\left( \mu, \sigma \right)}{P\left( \lbrace x_i \rbrace \right)} \, ,
\end{align*}

\noindent where $P\left( \mu, \sigma \vert \lbrace x_i \rbrace \right)$ is the joint posterior for $\mu$ and $\sigma$, $P\left(\lbrace x_i \rbrace \vert \mu, \sigma \right)$ is the sampling distribution or likelihood, $P\left( \mu, \sigma \right)$ is the joint prior distribution for $\mu$ and $\sigma$, and $P\left(\lbrace x_i \rbrace \right)$ acts as a normalizing constant.
The most important component is the sampling distribution $P\left( \lbrace x_i \rbrace \vert \mu, \sigma \right)$, as this acts as our model for the data.
As mentioned in the main text, we use a log-normal model for the dataset, such that

\begin{align*}
P\left( \lbrace x_i \rbrace \vert \mu, \sigma \right) &=\prod_{i=1}^N P\left( x_i \vert \mu, \sigma \right) \\ &= \prod_{i=1}^N \frac{1}{\sqrt{2 \pi \sigma^2}} \frac{1}{x_i} e^{-\sfrac{\left( \mathrm{ln}\! \left( x_i \right) - \mu \right)^2}{2 \sigma^2}} \, .
\end{align*}

For the prior distribution $P\left( \mu, \sigma \right)$ we choose uninformative priors.
Typical uninformative priors for $\mu$ and $\sigma$ are the uniform and Jeffreys priors, respectively\,\cite{sivia_data_2006}, but as they are improper distributions (i.e. not normalizable), it is common to start with proper (i.e. normalizable) distributions and take a limit at the end of the calculation to turn the proper priors into the desired uninformative priors\,\cite{gelman_bayesian_2014}.
As we can rewrite $P \left( \mu , \sigma \right) = P \left( \mu \right) P \left( \sigma \right)$ (assuming our prior states of knowledge for $\mu$ and $\sigma$ are independent), we have to choose two priors.
For $P\left( \mu \right)$, we choose a uniform prior:

\begin{align*}
P\left(\mu \right) = \frac{1}{2 \mu_{UB}} I_{-\mu_{UB} \leq \mu \leq \mu_{UB}} \, ,
\end{align*}

\noindent where $I$ is the indicator function and causes $P \left( \mu \right)$ to be non-zero only within the bounds set by $\mu_{UB}$.
In the limit $\mu_{UB} \rightarrow \infty$, this goes to a uniform distribution that allows all real values of $\mu$.
We take this limit after finding $P\left( \mu, \sigma \vert \lbrace x_i \rbrace \right)$.
For $P\left( \sigma \right)$ we choose the Jeffreys prior, which is a uniform distribution on a logarithmic scale and is commonly used for scale parameters such as the standard deviation:

\begin{align*}
P\left( \sigma \right) = \frac{1}{\mathrm{ln}\! \left( \sigma_{UB} \right) - \mathrm{ln}\! \left( \sigma_{LB} \right)} \frac{1}{\sigma} I_{\sigma_{LB} \leq \sigma \leq \sigma_{UB}} \, ,
\end{align*}

\noindent where $\sigma_{LB}$ and $\sigma_{UB}$ are the lower and upper bounds on $\sigma$.
We next calculate $P \left( \mu, \sigma \vert \lbrace x_i \rbrace \right)$ by combining our expressions for $P \left( \lbrace x_i \rbrace \vert \mu \sigma \right)$, $P \left( \mu \right)$, and $P \left( \sigma \right)$, and we also use the fact that $P\left(\lbrace x_i \rbrace \right) = \int_{\sigma_{LB}}^{\sigma_{UB}} \! \int_{-\mu_{UB}}^{\mu_{UB}} P\left(\lbrace x_i \rbrace \vert \mu, \sigma \right) P\left( \mu \right) P \left( \sigma \right) \, \mathrm{d}\mu \, \mathrm{d}\sigma$.
After taking the limits $\mu_{UB} \rightarrow \infty$, $\sigma_{LB} \rightarrow 0$, and $\sigma_{UB} \rightarrow \infty$, we find

\begin{widetext}
\begin{align}
P \left( \mu, \sigma \vert \lbrace x_i \rbrace \right) = \frac{2^{1-\frac{N}{2}} N^{\frac{N}{2}} \left( \overline{X^2} - \overline{X}^2 \right)^{\frac{N-1}{2}}}{\sqrt{\pi} \Gamma \left( \frac{N-1}{2} \right)} \left( \frac{1}{\sigma} \right)^{N+1} e^{\sfrac{-N \left( \overline{X^2} - 2 \overline{X} \mu + \mu^2 \right)}{2 \sigma^2}} I_{-\infty \leq \mu \leq \infty} I_{0 \leq \sigma \leq \infty} \, . \label{Eq6}
\end{align}
\end{widetext}

\noindent From here on, we leave out the indicator functions $I_{\infty \leq \mu \leq \infty}$ and $I_{0 \leq \sigma \leq \infty}$ for ease of notation, but they are always implicitly there. Note that this derivation relies on the assumption that $N \geq 2$, i.e. the dataset or data subset has at least two data points in it.

Now we can calculate the marginal posteriors for $\mu$ and $\sigma$, which summarize how much our data determine those parameters.
The marginal posterior for $\mu$ is defined as follows:

\begin{align*}
P\left( \mu \vert \lbrace x_i \rbrace \right) = \int_{0}^{\infty} \! P\left( \mu, \sigma \vert \lbrace x_i \rbrace \right) \, \mathrm{d}\sigma \, .
\end{align*}

\noindent Using our expression for $P\left( \mu, \sigma \vert \lbrace x_i \rbrace \right)$ in Eq.~\ref{Eq6}, we find

\begin{align*}
P&(\mu \vert \lbrace x_i \rbrace) = \nonumber \\ &\frac{\Gamma \left( \frac{\nu_\mu + 1}{2} \right)}{\Gamma \left( \frac{\nu_\mu}{2} \right) \sqrt{\pi \nu_\mu {\sigma_\mu}^2}} \left( 1 + \frac{1}{\nu_\mu} \left( \frac{\mu - \mu_\mu}{\sigma_\mu} \right)^2 \right)^{-\frac{\nu_\mu + 1}{2}} \, ,
\end{align*}

\noindent which is Eq.~\ref{Eq2} of the main text.

Similarly, $P\left( \sigma \vert \lbrace x_i \rbrace \right)$ is defined by

\begin{align*}
P\left( \sigma \vert \lbrace x_i \rbrace \right) = \int_{-\infty}^{\infty} \! P\left( \mu, \sigma \vert \lbrace x_i \rbrace \right) \, \mathrm{d}\mu \, ,
\end{align*}

\noindent which in our case yields

\begin{align*}
P\left( \sigma \vert \lbrace x_i \rbrace \right) = \frac{2 \beta_\sigma^{\alpha_\sigma}}{\Gamma\left( \alpha_\sigma \right)} \left( \frac{1}{\sigma} \right)^{2 \alpha_\sigma + 1} e^{\sfrac{-\beta_\sigma}{\sigma^2}} \, ,
\end{align*}

\noindent where $\alpha_\sigma$ and $\beta_\sigma$ are defined in the main text.
Note that this is a distribution for the standard deviation $\sigma$, not the variance $\sigma^2$.
To find the distribution for $\sigma^2$, we perform a change of variables and find

\begin{align*}
P(\sigma^2 \vert \lbrace x_i \rbrace) = \frac{{\beta_\sigma}^{\alpha_\sigma}}{\Gamma \left( \alpha_\sigma \right)} \left(\frac{1}{{\sigma}^2} \right)^{\left( \alpha_\sigma + 1 \right)} e^{-\sfrac{\beta_\sigma}{{\sigma}^2}} \, ,
\end{align*}

\noindent which is the Inverse Gamma distribution of Eq.~\ref{Eq3} in the main text.

Finally, we derive the posterior predictive distribution, which summarizes what the next data point could be, based on the data taken so far.
The posterior predictive distribution is defined as follows:

\begin{align*}
P\left( \tilde{x} \vert \lbrace x_i \rbrace \right) &= \nonumber \\ \int_{-\infty}^{\infty} &\int_{0}^{\infty} \! P\left(\tilde{x} \vert \mu, \sigma \right) P\left( \mu, \sigma \vert \lbrace x_i \rbrace \right) \, \mathrm{d}\sigma \mathrm{d}\mu \, ,
\end{align*}

\noindent which leads to Eq.~\ref{Eq5} of the main text:

\begin{align*}
P&(\tilde{x} \vert \lbrace x_i \rbrace) = \nonumber \\ &\frac{\Gamma \left( \frac{\tilde{\nu} + 1}{2} \right)}{\Gamma \left( \frac{\tilde{\nu}}{2} \right) \sqrt{\pi \tilde{\nu} \tilde{\sigma}^2}} \frac{1}{\tilde{x}} \left( 1 + \frac{1}{\tilde{\nu}} \left( \frac{ \mathrm{ln} \! \left( \tilde{x} \right) - \tilde{\mu} }{\tilde{\sigma}} \right)^2 \right)^{- \frac{\tilde{\nu} + 1}{2}} \, ,
\end{align*}

\noindent where $\tilde{\nu}$, $\tilde{\mu}$, and $\tilde{\sigma}$ are defined above.
Making a change of variables using $\tilde{X} = \mathrm{ln} \left( \tilde{x} \right)$ leads to Eq.~\ref{Eq4} of the main text:

\begin{align*}
P&(\tilde{X} \vert \lbrace x_i \rbrace) = \nonumber \\ &\frac{\Gamma \left( \frac{\tilde{\nu} + 1}{2} \right)}{\Gamma \left( \frac{\tilde{\nu}}{2} \right) \sqrt{\pi \tilde{\nu} \tilde{\sigma}^2}} \left( 1 + \frac{1}{\tilde{\nu}} \left( \frac{ \tilde{X} - \tilde{\mu} }{\tilde{\sigma}} \right)^2 \right)^{- \frac{\tilde{\nu} + 1}{2}} \, ,
\end{align*}

\noindent which is a location-scale t-distribution.

\section{Simulated Draws} \label{AppB}

Simulating draws is a common technique in Bayesian statistics to estimate credible intervals and answer probabilistic questions\,\cite{gelman_bayesian_2014}. To simulate draws from the posterior and posterior predictive distributions, we use Matlab's makedist function in the Statistics and Machine Learning Toolbox to define location-scale t-distributions and inverse gamma distributions with parameters determined by the data, as described in Section V.1.
Matlab's random function then allows us to sample from the distributions we defined based on our data.
Once samples have been drawn, it is simple to estimate the probabilities we describe above. For example, if we label our samples from $P \left( \mu_\mathrm{N14} \vert \lbrace x_i \rbrace \right)$ and $P \left( \mu_\mathrm{N15} \vert \lbrace x_i \rbrace \right)$ as $\lbrace \hat{\mu}_\mathrm{N14} \rbrace$ and $\lbrace \hat{\mu}_\mathrm{N15} \rbrace$, respectively, then we can estimate $P \left( \mu_\mathrm{N14} < \mu_\mathrm{N15} \vert \lbrace x_i \rbrace \right)$:

\begin{align*}
P \left( \mu_\mathrm{N14} < \mu_\mathrm{N15} \vert \lbrace x_i \rbrace \right) \approx \mathrm{mean} \left( \hat{\mu}_\mathrm{N14} < \hat{\mu}_\mathrm{N15} \right),
\end{align*}

\noindent where $\mathrm{mean} \left( \hat{x} \right)$ is the Matlab command for taking the mean of a vector\,\cite{gelman_bayesian_2014}.

\end{document}